\documentclass[12pt,preprint]{aastex}

\usepackage{emulateapj5}
\usepackage{onecolfloat}
\usepackage{graphicx}
\usepackage{times}

\newcommand{\sm}{\, {\rm M}_{\odot}}

\newcommand{\fgas}{f_{\rm gas}}
\def\gsim { \lower .75ex \hbox{$\sim$} \llap{\raise .27ex \hbox{$>$}} }
\def\lsim { \lower .75ex \hbox{$\sim$} \llap{\raise .27ex \hbox{$<$}} }

\newcommand{\cutt}[1]{}

\shorttitle{Dark satellites and dwarf galaxies}
\shortauthors{A. Helmi et al.}

\begin{document}

\twocolumn[

\title{Dark satellites and the morphology of dwarf galaxies}

\author{Amina Helmi\altaffilmark{1},  L.~V.~Sales\altaffilmark{2}, E.
  Starkenburg\altaffilmark{1,3}, T.~K.~Starkenburg\altaffilmark{1},
C.~A.~Vera-Ciro\altaffilmark{1}, G.~De Lucia\altaffilmark{4},
  Y.-S. Li\altaffilmark{5}}

\begin{abstract}

  One of the strongest predictions of the $\Lambda$CDM cosmological
  model is the presence of dark satellites orbiting all types of
  galaxies. We focus here on the dynamical effects of such satellites
  on disky dwarf galaxies, and demonstrate that these encounters can
  be dramatic. Although mergers with $M_{\rm sat} > M_{d}$ are
  not very common, because of the lower baryonic content they occur
  much more frequently on the dwarf scale than for $L_*$-galaxies. As
  an example, we present a numerical simulation of a 20\% (virial)
  mass ratio merger between a dark satellite and a disky dwarf (akin to
  the Fornax dwarf galaxy in luminosity) that shows that the merger
  remnant has a spheroidal morphology. We conclude that
  perturbations by dark satellites provide a plausible path for
  the formation of dSph systems and also could trigger 
  starbursts in gas rich dwarf
  galaxies. Therefore the transition from disky to the often amorphous, irregular,
  or spheroidal morphologies of dwarfs could be a natural consequence
  of the dynamical heating of hitherto unobservable dark
  satellites.
\end{abstract}

] 

\altaffiltext{1}{Kapteyn Astronomical Institute, University of Groningen,
P.O.Box 800, 9700 AV Groningen, The Netherlands.
{\sf{e-mail: ahelmi@astro.rug.nl}}}
\altaffiltext{2}{Max-Planck-Institut f\"{u}r Astrophysik,
Karl-Schwarzschild-Str. 1, D-85748, Garching, Germany}
\altaffiltext{3}{Department of Physics and Astronomy, University of Victoria,
  Victoria, BC V8P 5C2, Canada}
\altaffiltext{4}{INAF - Astronomical Observatory of Trieste, via G.B. Tiepolo 11, 
I-34143 Trieste, Italy}
\altaffiltext{5}{Kavli Institute for Astronomy and Astrophysics, Peking University, Beijing 100871, China}

\section{Introduction}
\label{sec:intro}

According to the $\Lambda$CDM scenario, stellar disks are immersed in
dark matter halos and are surrounded by a full spectrum of satellite
companions.  Encounters with these satellites can inject significant
amounts of energy into the system, with consequences that vary from
negligible to fully catastrophic disk destruction depending on the
relative mass of the perturber and the configuration of the event
(relative distances and velocities). Disk heating by such
substructures has been addressed in previous work
\citep{Toth1992,Quinn1993,Font2001,Benson2004}, but has generally
focused on the effect on bright Milky Way-like galaxies.
 
Cold dark matter models predict the structure of halos to be
self-similar; in such a way that, when properly scaled, a Milky
Way-sized halo looks comparable to one hosting a faint dwarf galaxy
\citep{Moore1999b,AqPaper2,Klimentowski2010,Wang2012}. However, galaxy
formation is not a self-similar process, as the properties of galaxies
depend in a complex way on e.g.\ the mass of their host halos. For
example, low mass (dwarf) galaxies are much more inefficient at
forming stars \citep{Blanton2001,RK2008} and have much higher
mass-to-light ratios than larger galaxies
\citep{Yang2003,walker2009}. In addition, gas cooling is likely to be
(nearly) completely inhibited in dark matter halos with masses below
$\sim 10^8 h^{-1} \sm$ \citep{Kaufmann2007}, which implies that the
satellites of dwarfs should be generally completely dark in contrast to
satellites in galaxy clusters or 
around $L_*$-galaxies.

In this  {\it Letter} we show that these considerations imply that the dynamical perturbations of
dark-matter satellites on dwarf galaxies are much more important than
on $L_*$-galaxies. Dark satellites may provide a channel for the
formation of dwarf spheroidal galaxies without the need to recur to
environmental effects \citep{Mayer2010} or multiple body interactions
\citep{Sales2007}.  Such interactions may also be responsible for the observed
increase of disk ``thickness'' towards fainter
galaxies \citep{Yoachim2006}, as well as explain the existence of
isolated dwarfs undergoing intense starbursts without an apparent
trigger \citep{Bergvall2011} as a result of a major merger with a dark
companion (T.~K.~Starkenburg et al., in prep.).

\section{Models}
\label{sec:model}

Our goal is to quantify the effects of substructures on disk-like
galaxies over a broader region of parameter space (and specifically mass
range) than done in previous work. To this end, we use the
second resolution level of the {\it Aquarius Simulations}
\citep{AqPaper2} and study the assembly history of {\it main} (as
opposed to satellite) dark matter halos in the mass range
$10^8-10^{12} h^{-1}M_\odot$. We follow their evolution from $t=2$ Gyr
onwards ($z \leq 3$), since by this time all halos in our sample have
accreted at least 10\% of their final mass, and the concept of
``main/host'' is well-defined.  At the present time, we find 739
such halos, including the six main {\it Aquarius} Milky
Way-like objects. We have identified substructures in these halos with
the SUBFIND algorithm \citep{Springel2001a} and tracked their orbits
by following the position of their most-bound particle.

We populate these dark matter halos with ``galaxies'' following a
semi-analytic model that uses simple but physically motivated laws to
track the evolution of gas cooling, star formation and feedback
processes \citep{Li2010,Else2011}. This allows us to derive their
baryonic properties such as their gas content, stellar mass, etc.  Our
model simultaneously reproduces the luminosity function, scaling
relations and chemical content of bright as well as dwarf galaxies
\citep[for a more detailed description see][]{Else2011}.

When a disk galaxy accretes a low mass companion, it is (vertically) heated
and puffed up. The 
increase in the scale height $\Delta H$ for a disk of (total, i.e. stellar and gas) mass $M_d$ and
scale length $R_d$ caused by an interaction with a satellite of
mass $M_{\rm sat}$ may be estimated using analytic arguments to be 
\begin{eqnarray}
\label{eq:H}
& & \frac{\Delta H}{R_d}  = \alpha (1-\fgas)\,\frac{M_{\rm
sat}}{M_d}, 
\end{eqnarray}
\citep{Toth1992,galform-book}. Here $\fgas=M_{\rm gas}/M_d$ is the gas
fraction of the host disk and its inclusion in Eq.~(\ref{eq:H})
accounts for the energy that is radiated away and not transferred into
random motions of disk stars \citep[e.g.][]{Hopkins2008}. We have
carried out a series of merger experiments on the scale of dwarfs
\citep[and used analogous simulations of large
disks by][]{velazquez1999,vh2008,purcell,moster}, and confirm
the above dependence on the ratio $M_{\rm sat}/M_d$. We have found the
proportionality constant to be $\alpha \sim 0.03$ when the above
expression is evaluated at $R = 2.5 R_d$. Below we present two
examples of such merger simulations and report our results in more detail in
T.~K.~Starkenburg et al., in prep.
\begin{figure*}
\includegraphics[height=0.35\textheight,clip]{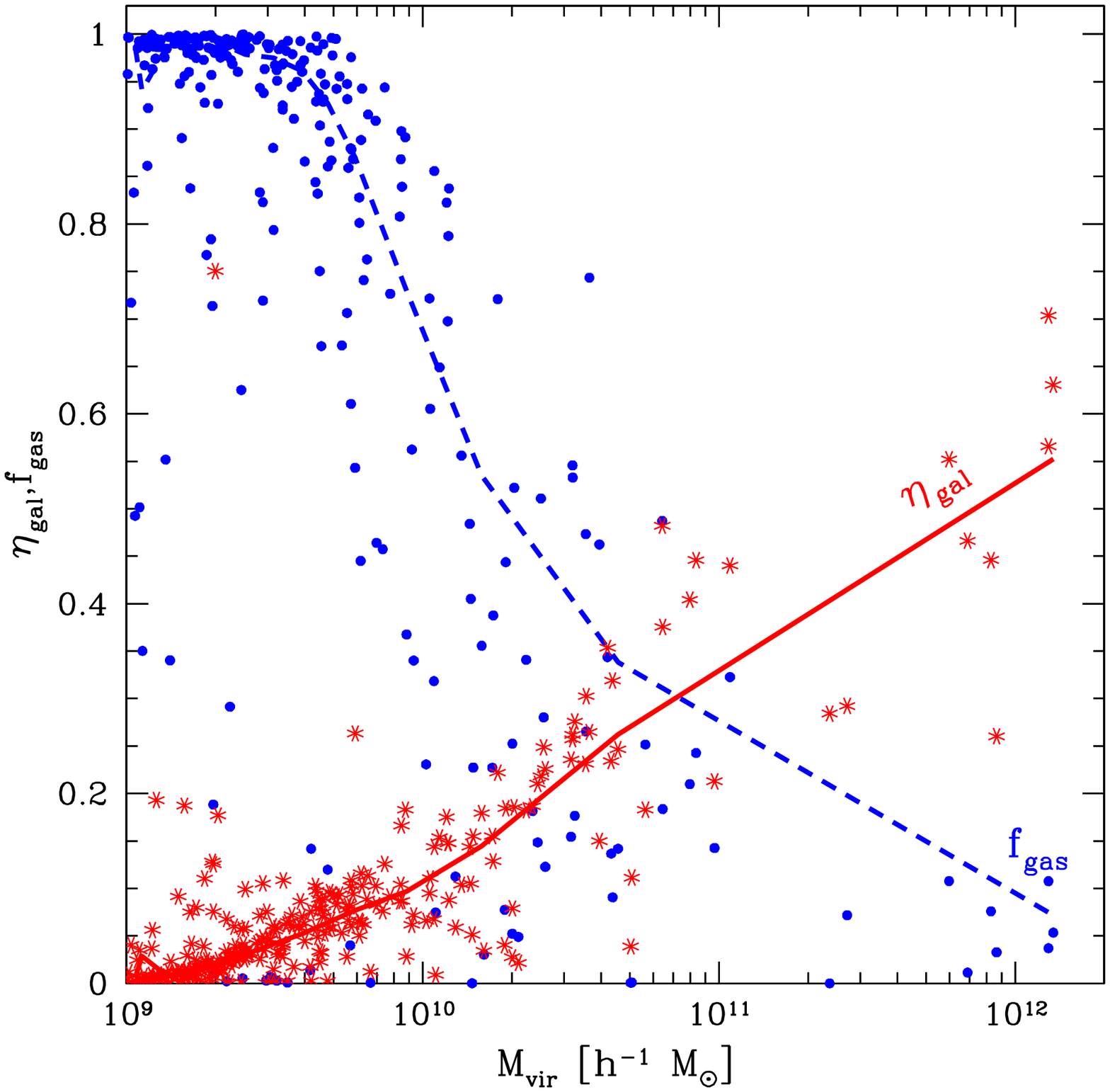}
\includegraphics[height=0.35\textheight,clip]{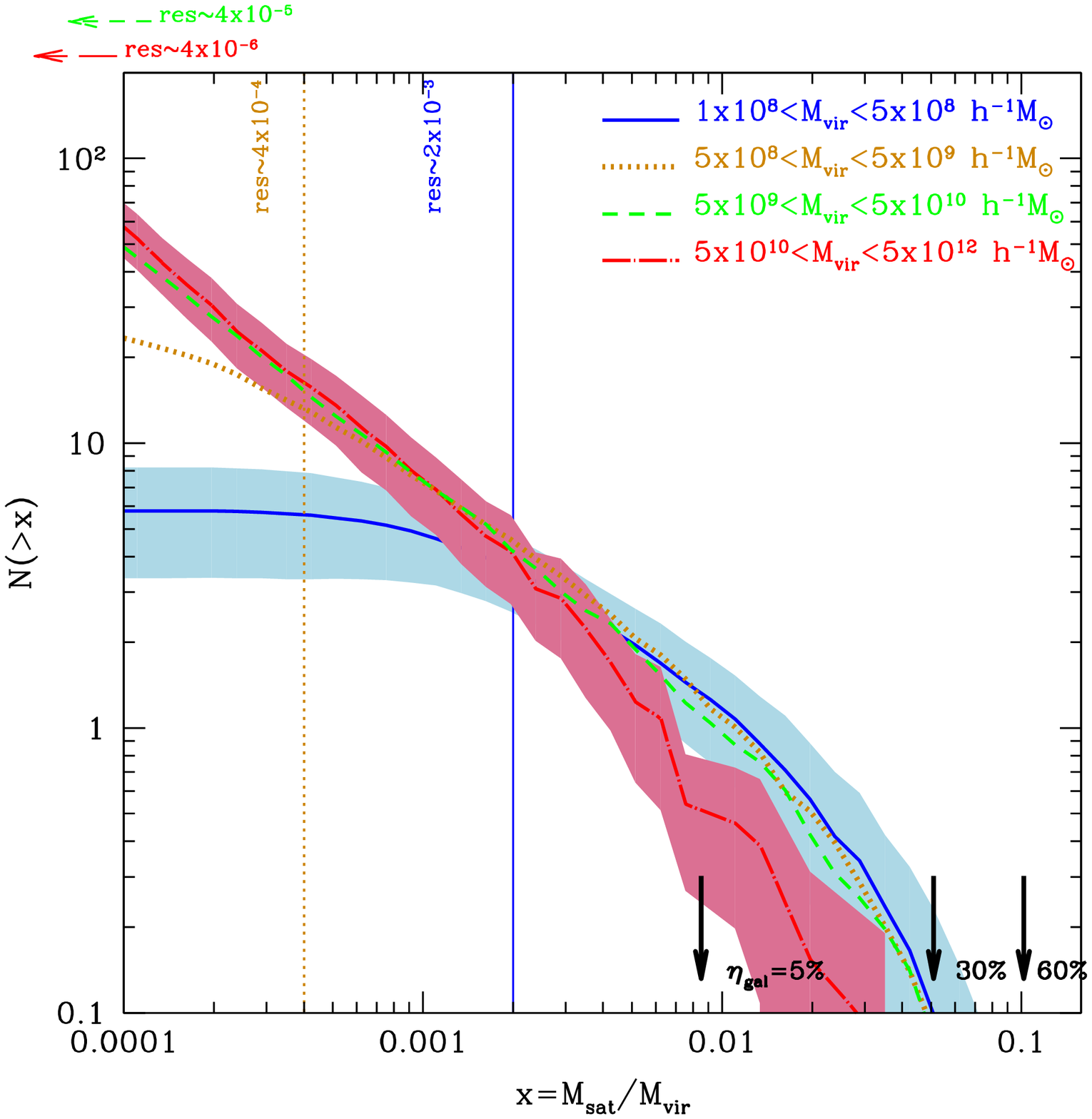}
\caption{{\it Left:} Gas fraction $\fgas$ (blue circles) and galaxy
  formation efficiency $\eta_{\rm gal}$ (red asterisks) as a function of host mass as predicted by our
 SA model. The dashed blue and solid red curves indicate
  the respective median trends. {\it Right:} The ``Spectrum of
  perturbers'' $N (> x)$ gives the number of encounters with
  objects of a given mass ratio $x = M_{\rm sat}/M_{\rm vir}$ (see
  text for more details). The shaded regions correspond to the 25 and
  75 percentiles and were derived from 100 random subsamples with each
  5 host halos belonging to a given mass range
  (and are shown for clarity only for the two mass ranges). The black arrows
  indicate $M_d/M_{\rm vir}$ for three representative values of
  $\eta_{\rm gal}=5, 30$ and $60\%$, while the thin vertical lines are the
  SUBFIND mass resolution.}
\label{fig:fig1}
\end{figure*}

Eq.~(\ref{eq:H}) can be re-written in terms of the ``disk galaxy efficiency''
of a given halo: $\eta_{\rm
gal}=M_d/(M_{\rm vir}\times f_{\rm bar})$, i.e.\ the fraction of baryons collected in the
central galaxy compared to the total available budget. Here $M_{\rm
  vir}$ is the
virial mass of the host halo and $f_{\rm bar}\sim 0.17$ is the
universal baryon fraction. Therefore
\begin{equation}
\frac{\Delta H}{R_d} = \frac{\alpha}{f_{\rm bar}}\, \frac{(1-\fgas)}{\eta_{\rm gal}}\,
\frac{M_{\rm sat}}{M_{\rm vir}}.
\label{eq:H2}
\end{equation}
Thus three quantities affect the efficiency of disk
heating: the gas fraction $\fgas$, the galaxy efficiency $\eta_{\rm
  gal}$ and the mass of the perturber compared to that of the host
$M_{\rm sat}/M_{\rm vir}$. We now investigate each of these factors
using our models.

The blue solid circles in the left panel of Fig.~\ref{fig:fig1} show
$\fgas$ as a function of host halo mass in the SA model. Note that the gas content of
a galaxy depends strongly on the mass of its halo: for objects less
massive than $10^{10} h^{-1}M_\odot$ more than 90\% of the baryonic
mass assembled onto the central galaxy remains as cold gas, revealing
how inefficient star formation is in (isolated) dwarf
galaxies. On the other hand, Milky Way-sized objects have typically
$\sim 10-20 \%$ of their baryons in gas; all these numbers being in
reasonably good agreement with observations \citep{McGaugh2010}.

The red asterisks in the left panel of Fig.~\ref{fig:fig1} show
$\eta_{\rm gal}$ as function of halo mass. As indicated by the median
trend (red solid line), halos become increasingly inefficient in
collecting baryons onto galaxies as they become less massive: for
$M_{\rm vir}< 10^{10} h^{-1} M_\odot$, $\eta_{\rm gal} \sim
1-10\%$. This is the result of a combination of the effect of a UV
ionizing background and of supernova feedback
\citep{Li2010,Maccio2010,Okamoto2010}. These processes need to be
taken into account to match the satellite luminosity function
\citep{Guo2010,Moster2010}, and explain why dwarf galaxies are the
most dark matter dominated objects known in the Universe.

The right panel of Fig.~\ref{fig:fig1} shows the cumulative subhalo
mass function for our sample of main halos in the {\it
  Aquarius} simulations for four different ranges of host mass. Since
disk heating is expected to be more efficient for perturbers that
adventure close to the center of the host halo, we measure the subhalo
mass at the first pericenter that is within a distance smaller than
30\% of the virial radius of the host and normalize it to the
virial mass of the host at that time. The thin vertical lines indicate
the subhalo resolution, defined by the 20-particle threshold imposed
by the SUBFIND algorithm\footnote{Note that for the least massive host
  halos we are able to resolve fewer substructures than for Milky
  Way-like hosts, and that the SUBFIND algorithm is known to
  underestimate the mass at pericenter, hence the above values are
  lower limits.}.  Within the range that is well resolved (to the
right of the vertical lines), we find that the mass spectra of satellites at
pericenter are all comparable and independent of the virial mass of
the host.

Because the efficiency of galaxy formation $\eta_{\rm gal}$ depends
strongly on $M_{\rm vir}$ (see left panel of Fig. \ref{fig:fig1}), at fixed gas content the
heating produced by satellites is expected to be significantly larger
for small mass hosts (halos with $M_{\rm vir}<10^{10} h^{-1} M_\odot$) than
for Milky Way-like galaxies. To first order this is hinted by the
vertical arrows in the right panel of Fig.~\ref{fig:fig1}. These
arrows show the mass ratio $M_{d}/M_{\rm vir}$ for three different
values of galaxy efficiency: $\eta_{\rm gal}=5,30$ and $60\%$, and can
be used as a guide to determine the number of encounters with
satellites with $M_{\rm sat} \sim M_d $ for a system with a given
efficiency. It then becomes clear that such encounters are much more
common for dwarf galaxies, which have lower $\eta_{\rm gal}$. For
example, a dwarf galaxy ($\eta_{\rm gal}\sim 5\%$), experienced on
average 1.5 encounters with an object of comparable mass in the last
11.7 Gyr. On the other hand, for a Milky Way-like galaxy whose disk
mass is $\sim 10\%$ of the virial value ($\sim 5\%$ for $\eta_{\rm
  gal}=30\%$), the number of encounters with a significant perturber
are a factor $\sim 15$ less common.  Note that these estimates are
somewhat lower than derived in previous work for $\sim 10^{12} \sm$
hosts \citep{purcell} and
this could be due to the environment of the {\it Aquarius}
halos.

\section{Results}

\begin{figure}
\begin{center}
\includegraphics[height=0.35\textheight,clip]{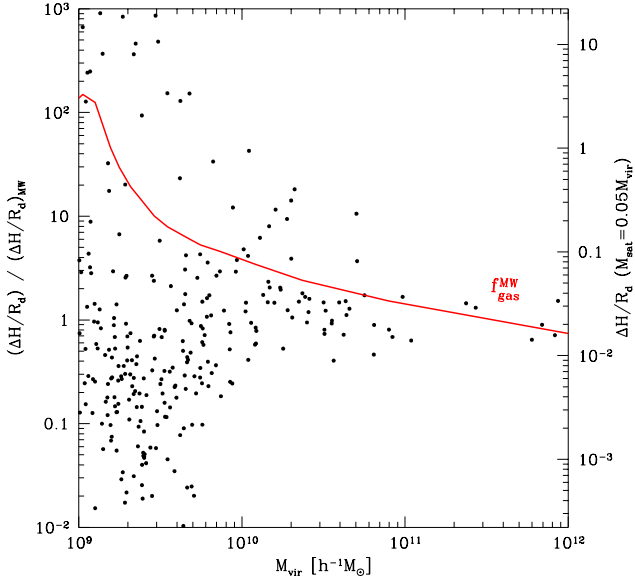}
\end{center}
\caption{Relative increase in the disk's thickness $\Delta
  H/R_d$ as a function of the virial mass of the host computed using
  Eq.~(\ref{eq:H2}) for all the main galaxies in the Aquarius simulations
  (black dots). The red curve shows the median expected change for $f_{\rm
    gas} = 0.1$. All values have been normalized to the
  heating expected for an $L_*$ galaxy, with $f_{\rm gas} = 0.1$ and
  $\eta_{\rm gal} = 0.45$.  The scale on the right vertical axis
  indicates the absolute change in $\Delta H/R_d$ assuming an
  encounter with a $M_{\rm sat}= 0.05 M_{\rm vir}$ perturber.
\label{fig:fig2}}
\end{figure} 

\begin{figure*}
\begin{center}
\includegraphics[height=0.2\textheight,clip]{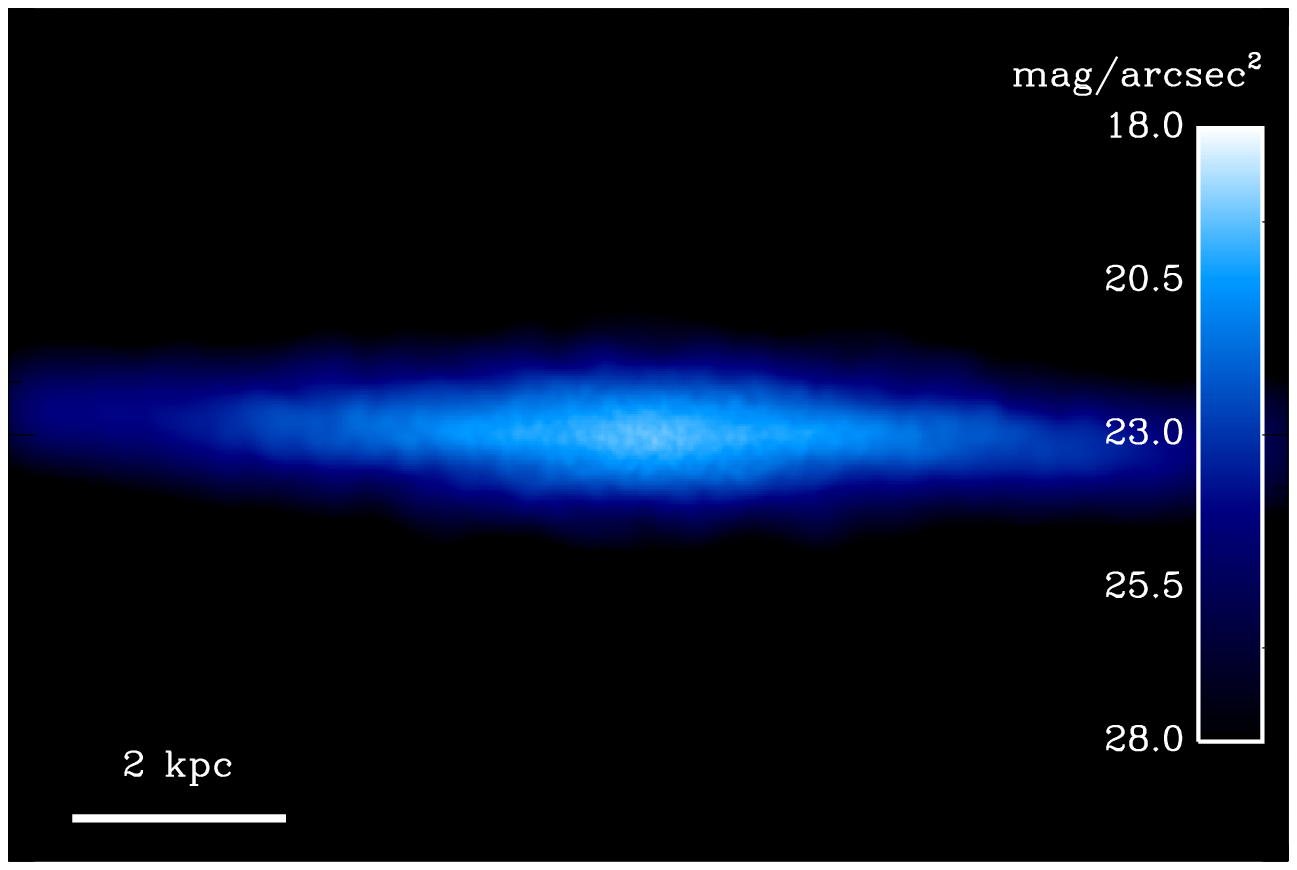}
\includegraphics[height=0.2\textheight,clip]{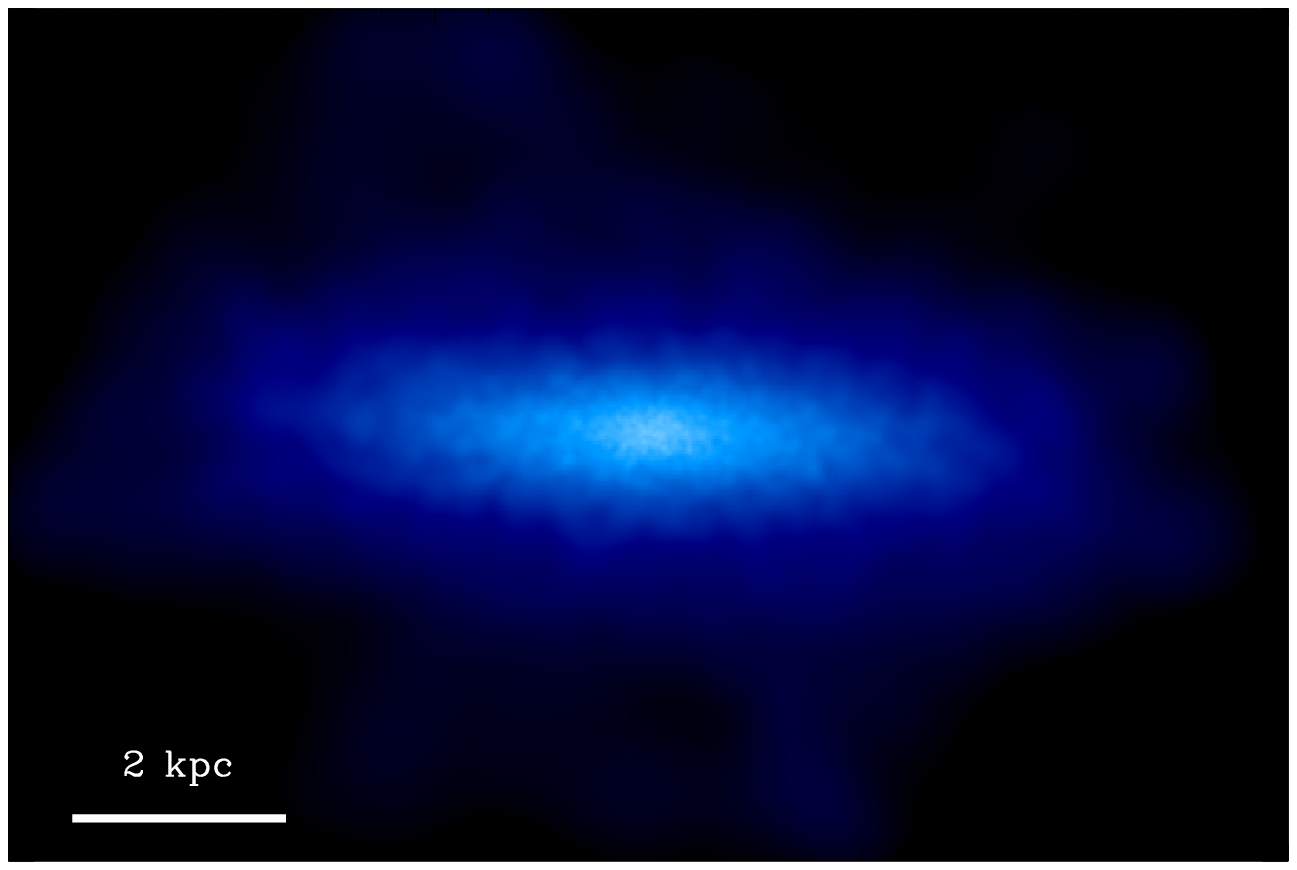}
\includegraphics[height=0.2\textheight,clip]{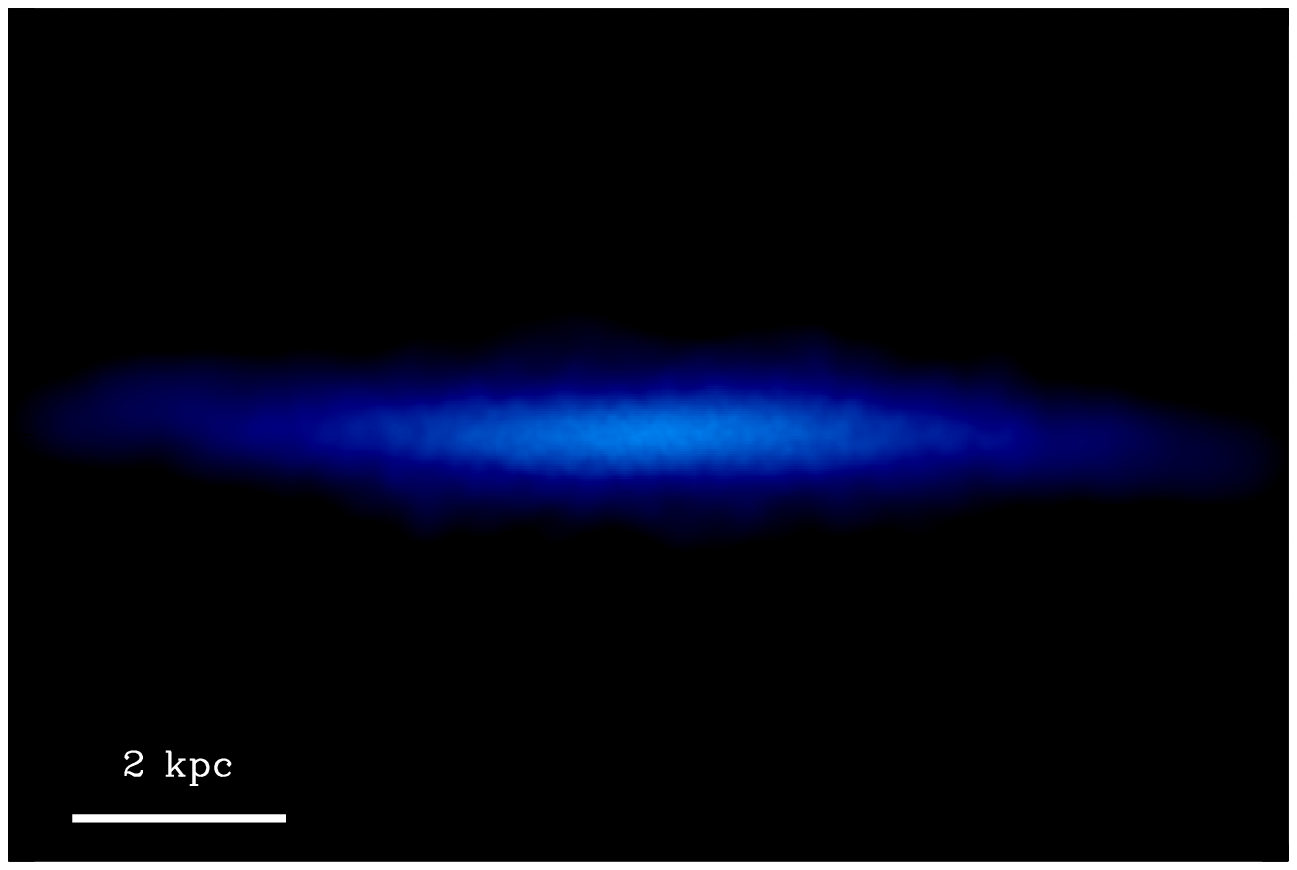}
\includegraphics[height=0.2\textheight,clip]{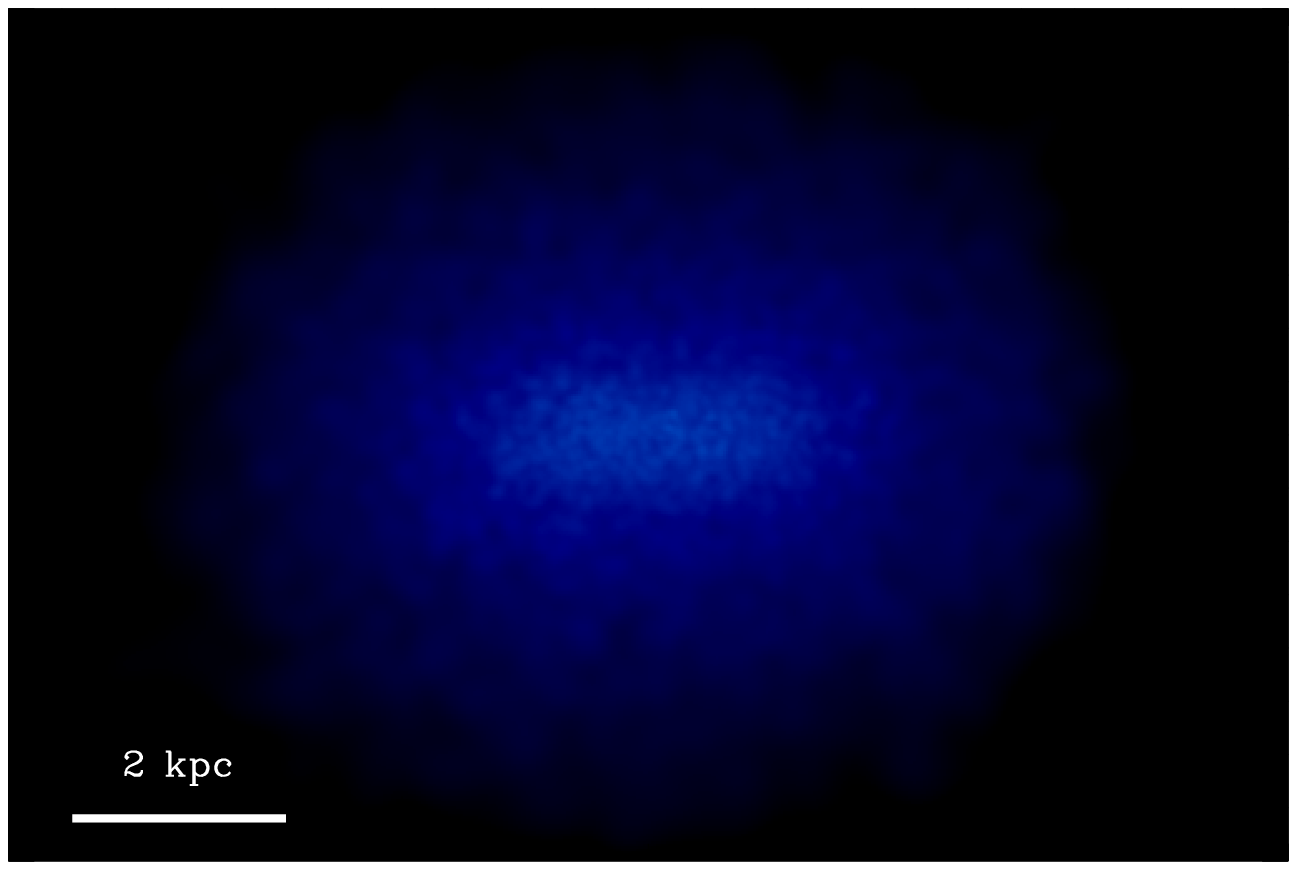}
\end{center}
\caption{The left panels show the initial surface brightness profiles
  for two of our simulated dwarf galaxies with $M_d/M_{\rm vir} = 0.008$ and
  $0.04$ (bottom and top panels, respectively). The
  panels on the right correspond to the final stellar distributions
  after these disks merged with a dark satellite of mass $M_{\rm sat}
  = 0.2 M_{\rm vir}$, and are shown after 6 Gyr of evolution (i.e well
  after the merger has taken place, so the system appears to be
  relaxed again).}
\label{fig:surf_brightness}
\end{figure*} 

Fig.~\ref{fig:fig2} shows, for our model galaxies, $\Delta H/R_d$ as
function of host halo mass, normalized to the values expected for a
galaxy like the Milky Way (with $\eta_{\rm gal} = 0.45$ and $f_{\rm gas} =
0.1$) and for fixed $M_{\rm sat}/M_{\rm vir}$. 
The red curve indicates the expected change {\it when the gas fraction is
that of the Milky Way}. This shows that for a dwarf galaxy populating a
$10^9 h^{-1}M_\odot$ halo, the heating of a disk is expected to be
$\sim 100$ times larger than for a galaxy like the Milky Way
embedded in a $10^{12} h^{-1}M_\odot$ halo.  For example, even an
encounter with a low mass perturber ($M_{\rm sat}/M_{\rm vir}=0.05$)
would be devastating and turn a disky dwarf galaxy into a dwarf
spheroidal since $\Delta H/R_d \sim 2.7$ for $M_{\rm vir} = 10^9
h^{-1}M_\odot$ and $f_{\rm gas} = 0.1$ according to Eq.~(\ref{eq:H2}). On the other hand, the effect
of such an encounter would be nearly negligible in the case of a
Milky-Way like galaxy.

On the scale of Milky Way galaxies the heating is dominated by
subhalos hosting stars, while around smaller
hosts ($M_{\rm vir}<10^{10} h^{-1}M_\odot$) the subhalos will generally be {\it
  dark} as they fall below the mass threshold imposed by reionization
and efficient atomic hydrogen cooling to form stars.
To confirm that such dark satellites leave imprints on the
morphologies of dwarf galaxies, we have performed a set of numerical
experiments. We focus here on two simulations where we varied the mass
ratio between the disk and the satellite, but took $M_{\rm sat}/M_{\rm
  vir} = 0.2$ comparable to what has been used in previous work
\citep{Kazantzidis2008,vh2008,purcell,Moster2010}.  The satellite
follows an NFW profile with concentration
$c = 18.7$  \citep{munoz2011}.  Our disk galaxies are purely stellar, they
have $M_d = 0.008$ and $0.04 \times M_{\rm vir}$, and are embedded in
a Hernquist halo with mass $M_{\rm vir} = 10^{10} h^{-1} \sm $,
and scale-radius $a = 9.3$~kpc,  i.e. $\eta_{\rm
  gal}\sim 5\%$ and $23\%$ respectively.  The disks are radially
exponential with scale-length $R_d = 0.67 h^{-1}$ kpc, and vertically
they follow ${\rm sech}^2(z/2z_0)$, with $z_0 = 0.05R_d$. The
internal kinematics are set-up following \citet{hernquist1993}, and
the disks are stable (with Toomre parameters $Q >1 $).

We put the satellite on a fairly radial orbit with $r_{\rm
  apo}/r_{\rm peri} =40$, starting from a distance of $\sim 23$ kpc,
and found that it is completely disrupted after three close passages,
i.e.  in $\sim 1.5$ Gyr. Fig.~\ref{fig:surf_brightness} shows the
final surface brightness profiles of the heavy and light disks in the
top and bottom panels respectively. This figure evidences that
significant heating has taken place and even led to important changes
in the morphology of the host galaxy. This is expected since although
we simulated minor mergers in terms of virial mass ratios, these are
major mergers from the perspective of the dwarf galaxy, as $M_{\rm
  sat}/M_d = 5$ and 25 respectively.

In the case of gas-rich systems --which are a majority at the low mass end, encounters with dark satellites will
be less efficient at changing the structure of the host dwarf galaxy,
because much of the orbital energy will be absorbed by the gas,
leading to less vertical heating. However, we may expect that such
encounters may induce star formation events, and thus, albeit
indirectly, lead to significant changes in the characteristics of
these galaxies (T.~K.~Starkenburg et al., in prep).

\begin{figure}
\begin{center}
\includegraphics[height=0.35\textheight,clip]{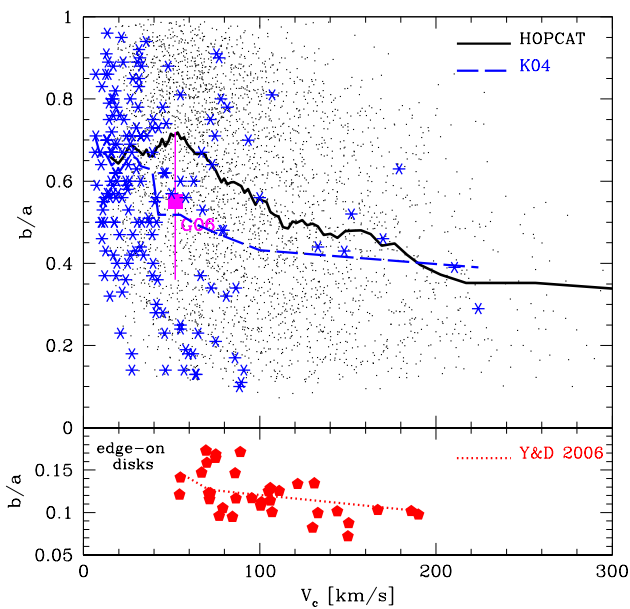}
\end{center}
\caption{Apparent optical axis ratios $b/a$ for various samples
of isolated, late-type galaxies from the literature.  {\it Top panel}: HOPCAT (Doyle et al. 2005;
  black dots), a sample of nearby galaxies selected from Karachentsev
  et al. (2004; blue asterisks) and the median of the sample of dwarf
  galaxies presented by Geha et al. (2006; magenta square with vertical
  bar showing the 25-75\% percentiles). {\it Bottom
    panel:} Edge-on galaxies from Yoachim \& Dalcanton (2006). These are
  shown separately because the $b/a$ corresponds to their true axis
  ratio, i.e. $z_0/R_d$. The lines indicate, for each sample, the median $b/a$ at a
  given circular velocity $V_c$.  }
\label{fig:fig4}
\end{figure}

To establish whether observations support that disks of dwarf galaxies are thicker
than those of larger systems, we have compiled measurements of the
thickness of {\it stellar} disks (quantified by the apparent axis
ratio, $b/a$) for a wide range of galaxy masses. Although the observed
$b/a$ is not a measurement of the intrinsic shape of the disk, if one
assumes random orientations on the sky, the two are directly
related. In our literature search we have carefully selected {\it
  isolated late-type} galaxies to avoid any morphology-luminosity
trend that may be driven by environmental interactions \citep[such as
discussed in][]{Mayer2010}.

The top panel of Figure \ref{fig:fig4} shows the distribution of
optical ($r$ or $R$ band) $b/a$ as a function of circular velocity
($V_{\rm c} = W_{50}/2$). The latter provides a measure of the
dynamical mass of the galaxy and its dark matter halo. We plot here
data for two galaxy samples: HOPCAT \citep[black dots;][]{Doyle2005} 
containing the optical counterparts of $\sim 3600$ HIPASS sources, and
for a set of isolated nearby late type galaxies
\citep[blue asterisks;][tidal index $\Theta<0$, RC3 morphological
type $>0$]{Karachentsev2004}. The magenta square shows the median value
for a subsample of the 101 dwarf galaxies \citep{Geha2006}, where we
have selected only those objects with no companions within $1$~Mpc
projected distance and $(g-r)<0.55$.  The bottom panel of this Figure
shows the intrinsic thickness defined as the ratio of scale-height to
scale-length for a sample of edge-on disks \citep{Yoachim2006}.

The median $b/a$ trends with galaxy circular velocity are indicated
separately for each sample by the black solid (HOPCAT), blue dashed
\citep{Karachentsev2004} and red dotted \citep{Yoachim2006}
curves. Each set clearly shows that the axis ratios increase with
decreasing circular velocities. In other words, stellar disks become
thicker as we move towards less massive galaxies, in
qualitative agreement with our expectations based on the analysis of disk
heating by substructure on isolated galaxies \citep[see also][]{Sanchez-Janssen2010}. 

\section{Discussion}

We have demonstrated that the dynamical effects of dark satellites on
disky dwarf galaxies are much more dramatic than on galaxies like the
Milky Way. Mergers with $M_{\rm sat} > M_{d}$ are not very common for
$z < 3$ but they occur much more frequently than on the $L_*$-galaxies
scale. As an example, we have simulated a merger with $M_{\rm
  sat}/M_{\rm vir} = 0.2$ for a dwarf with $M_d = 8 \times 10^{7}
h^{-1} \sm$ in stars, i.e. slightly more massive than the Fornax dwarf
galaxy, and found that its morphology changed from disky to
spheroidal.  This might be a plausible path for the formation of dSph
systems in isolation (if the dwarf was gas poor, which is rare in our
models but not unlikely). This channel might also be relevant for the
dSph satellites of our Galaxy, provided such encounters would have taken
place just before the system fell onto the potential well
of the Milky Way (since further gas accretion would thus be prevented).

Most of the galaxies on the scales of dwarfs are, however, gas-rich. In
that case, encounters with dark satellites can
trigger starbursts, which might explain the presence of seemingly
isolated dwarfs undergoing major star formation events without an
apparent trigger. Depending on the characteristics of the encounter,
such starbursts will vary in amplitude. We are currently performing hydrodynamical simulations to
characterize this process (T.~K.~Starkenburg et al. in prep).

Additionally, other processes exist that can influence the
morphologies of dwarf galaxies.  For example, binary mergers between
disky dwarfs can result in the formation of spheroidal systems
\citep{stelios2011}, although such events are rare (see Fig.~1). On
the other hand, on the scale of dwarf galaxies physical processes affecting gas may also
lead to thicker systems. For example,
the presence of a temperature floor in the interstellar medium at
$T\sim 10^4K$ introduced by e.g. a UV background, implies that gas
pressure becomes comparable to rotational support for small dark
matter halos. Stars formed in such systems would thus be born in
puffier configurations as demonstrated by \citet{Kaufmann2007}
\citep[see also, e.g.][]{RK2008}.

Yet, we have shown here that a distinctive imprint on dwarf galaxies
will be left by dark satellites in the context of the $\Lambda$CDM
cosmological paradigm. Such {\it dark satellites} are expected to make
the stellar disks of isolated dwarf galaxies significantly thicker
than those of $\sim L_*$ galaxies. We have indeed detected such a trend
on three different observational samples of isolated late-type
galaxies on the nearby Universe. We may have identified a
new mechanism to explain the morphologies of dwarf galaxies.

\acknowledgements
We are grateful to Volker Springel, Simon White and Carlos Frenk for
their generosity with respect to the use of the Aquarius simulations.
We thank Marla Geha for making her data available in electronic form.
ES, LVS and AH gratefully acknowledge financial support from NWO and
from the European Research Council under ERC-Starting Grant
GALACTICA-240271.  ES is supported by the Canadian Institute for
Advanced Research (CIfAR) Junior Academy and a Canadian Institute for
Theoretical Astrophysics (CITA) National Fellowship. LVS is grateful
for financial support from the CosmoComp/Marie Curie network. GDL
acknowledges financial support from the European Research Council
under the European Community's Seventh Framework Programme
(FP7/2007-2013) ERC grant agreement n. 202781.

\end{document}